\documentclass[aoas,MSNbibl,nameyear,dvips]{arximspdf}
\usepackage{dcolumn}
\usepackage{graphicx}

\doi{10.1214/12-AOAS616} 
\volume{7}
\issue{2}
\pubyear{2013}
\firstpage{640}
\lastpage{668}

\makeatletter

\newcolumntype{d}[1]{D{.}{.}{#1}}

\newcommand{\bbbeta}{\bolds{\eta}}
\newcommand{\bmu}{\bolds{\mu}}
\newcommand{\bsigma}{\bolds{\sigma}}
\newcommand{\bgamma}{\bolds{\gamma}}
\newcommand{\bTheta}{\bolds{\Theta}}
\newcommand{\bs}{\mathbf{s}}
\newcommand{\bc}{\mathbf{c}}
\newcommand{\bg}{\mathbf{g}}
\newcommand{\bT}{\mathbf{T}}

\makeatother

\begin{document}
\begin{frontmatter}

\title{Bayesian object classification of gold nanoparticles}
\runtitle{Bayesian object classification of gold nanoparticles}

\begin{aug}
\author[A]{\fnms{Bledar~A.}~\snm{Konomi}\corref{}\thanksref{t2,t3}\ead[label=e1]{alexandros@stat.tamu.edu}},
\author[A]{\fnms{Soma~S.}~\snm{Dhavala}\thanksref{t2,t3}\ead[label=e2]{soma@stat.tamu.edu}},
\author[A]{\fnms{Jianhua~Z.}~\snm{Huang}\thanksref{t1,t4}\ead[label=e3]{jianhua@stat.tamu.edu}},
\author[C]{\fnms{Subrata}~\snm{Kundu}\thanksref{t2}\ead[label=e4]{skundu@tamu.edu}},
\author[C]{\fnms{David}~\snm{Huitink}\thanksref{t2}\ead[label=e5]{dhuitink@tamu.edu}},
\author[C]{\fnms{Hong}~\snm{Liang}\thanksref{t2,t1}\ead[label=e6]{hliang@tamu.edu}},
\author[B]{\fnms{Yu}~\snm{Ding}\thanksref{t2,t5}\ead[label=e7]{yuding@iemail.tamu.edu}}
\and
\author[A]{\fnms{Bani~K.}~\snm{Mallick}\thanksref{t2,t3,t1}\ead[label=e8]{bmallick@stat.tamu.edu}}
\runauthor{B. A. Konomi et al.}
\affiliation{Texas A\&M University}
\address[A]{B. A. Konomi\\
S. S. Dhavala\\
J. Z. Huang\\
B. K. Mallick\\
Department of Statistics\\
Texas A\&M University\\
College Station, Texas 77843-3143\\
USA\\
\printead{e1}\\
\hphantom{E-mail: }\printead*{e2}\\
\hphantom{E-mail: }\printead*{e3}\\
\hphantom{E-mail: }\printead*{e8}} 
\address[B]{Y. Ding\\
Department of Industrial\\
\quad and Systems Engineering\\
Texas A\&M University\\
College Station, Texas 77843-3131\\
USA\\
\printead{e7}}
\address[C]{S. Kundu\\
D. Huitink\\
H. Liang\\
Department of Mechanical Engineering\\
Texas A\&M University\\
College Station, Texas 77843-3123\\
USA\\
\printead{e4}\\
\hphantom{E-mail: }\printead*{e5}\\
\hphantom{E-mail: }\printead*{e6}}
\end{aug}

\thankstext{t2}{Supported in part by the Texas Norman Hackerman
Advanced Research Program under Grant 010366-0024-2007.}

\thankstext{t3}{Supported in part by NSF Grant DMS-09-14951.}

\thankstext{t1}{Supported in part by King Abdullah University of
Science and Technology, Award Number KUS-CI-016-04.}

\thankstext{t4}{Supported in part by NSF Grants DMS-09-07170,
DMS-10-07618 and DMS-12-08952.}

\thankstext{t5}{Supported in part by NSF Grant CMMI-1000088.}

\received{\smonth{3} \syear{2011}}
\revised{\smonth{11} \syear{2012}}

%
\begin{abstract}
The properties of materials synthesized with nanoparticles (NPs) are
highly correlated to the sizes and shapes of the nanoparticles. The
transmission electron microscopy (TEM) imaging technique can be used to
measure the morphological characteristics of NPs, which can be simple
circles or more complex irregular polygons with varying degrees of
scales and sizes. A major difficulty in analyzing the TEM images is the
overlapping of objects, having different morphological properties with
no specific information about the number of objects present.
Furthermore, the objects lying along the boundary render automated
image analysis much more difficult. To overcome these challenges, we
propose a Bayesian method based on the marked-point process
representation of the objects. We derive models, both for the marks
which parameterize the morphological aspects and the points which
determine the location of the objects. The proposed model is an
automatic image segmentation and classification procedure, which
simultaneously detects the boundaries and classifies the NPs into one
of the predetermined shape families. We execute the inference by
sampling the posterior distribution using Markov chain Monte Carlo
(MCMC) since the posterior is doubly intractable. We apply our novel
method to several TEM imaging samples of gold NPs, producing the needed
statistical characterization of their morphology.
\end{abstract}

%

\end{frontmatter}

\section{Introduction}\label{secseca}

Nanoparticles (NPs) are tiny particles of matter with diameters
typically ranging from a few nanometers to a few hundred nanometers
which possess distinctive properties. These particles, larger than
typical molecules
but too small to be considered bulk solids, can exhibit hybrid physical
and chemical
properties which are absent in the corresponding bulk material. The
particles in their nano regime exhibit special properties which are not
found in the bulk properties, for example, catalysis [\citet
{Kundu2003}], electronic properties [\citet{Jana99b}] and size and
shape dependent optical properties [\citet{Jana99a}], which have
potential ramifications in medicinal applications and optical devices
[\citet{Link99,Kamat93}]. The current challenge is to develop
capabilities to understand and synthesize materials at the nano stage,
instead of the bulk stage.

Among the various NPs studied, colloidal gold (Au) NPs were found to
have tremendous importance due to their unique optical, electronic and
molecular-recognition properties
[\citet{Hirsch03} and \citet{Gaponik00}]. For example, selective
optical filters, bio-sensors, are among the many applications that use
optical properties of
gold NPs related to surface plasmon resonances which depend strongly on
the particle shape and size [\citet{Yu97}].
Moreover, there is an enormous interest in exploiting gold NPs in
various biomedical applications since their scale is similar to that of
biological molecules
(e.g., proteins, DNA) and structures (e.g., viruses and bacteria) [\citet
{Chitrani2006}].

In recent years it has become possible to investigate the dependency of
chemical and physical properties on size and shape of NPs, due to
Transmission Electron Microscopy (TEM) images. \citet{Sau2001} and
\citet{Kundu09b}, respectively, showed size and shape dependence of
synthesis and catalysis reaction where they observed different rates.
They also observed that circular gold NPs are better catalysts compared
to triangular NPs for a specific reaction. The development of new
pathways for the systematic manipulation of size and shape over
different dimensions is thus critical for obtaining optimal properties
of these materials. In this paper we develop novel, model-based image
analysis tools that classify and characterize the images of the NPs
which provide their morphological characteristics to enable a better
understanding of the underlying physical and chemical properties. Once
we are able to accurately characterize the shapes of NPs by using this
method,
we can develop different techniques to control these shapes to extract
useful material properties.

Substantial work in estimating the closed contours of objects in an
image has been done by \citet{Blake92,Qiang09,Pievatolo98,Jung08,Kothari09}, among others. Imaging processing tools, especially for cell
segmentation, also exist; for instance, ImageJ [\citet{ImageJ}] is a
tool recommended by the National Institute of Health (NIH). However,
the features of the data we are dealing with are quite different from
those considered in the literature reviewed, as there are various
degrees of overlapping of the NPs differing in shapes and sizes, as
well as a significant number of NPs lying along the image boundaries.

High-level statistical image analysis techniques model an image as a
collection of discrete objects and are used for object recognition
[\citet{Baddeley93}]. In images with object overlapping, Bayesian
approaches have
been preferred over maximum likelihood estimators (MLE). The unrestricted
MLE approaches tend to contain clusters of identical objects allowing
one object to
sit on the top of the other, whereas the
Bayesian approaches mitigate this problem by penalizing the overlapping as
part of the prior specification [\citet{Ripley77b,Baddeley93}],
offering flexibility over controlling the overlapping or
the touching.

In \citet{Mardia97}, a Bayesian approach using a prior which forbids
objects to overlap completely is proposed to capture predetermined shapes
(mushrooms, circular in shape). Inference is carried out by finding the
Maximum A Posteriori (MAP) estimates and the prior parameters are
chosen by
simulation experience, in effect, fixing the parameters that define the
penalty terms. \citet{Rue99} also used a similar framework to handle
the unknown number of objects but introduce polygonal templates to model
the objects. However, their application is restricted to cell detection
problems, where the objects do not overlap but barely touch each other and
the method works more like a segmentation technique than as a
classification technique. Moreover, the success of this approach
depends on
prior parameters, which are assumed known throughout the simulation.
\citet{Al2} used the same model except
that they considered elliptical templates instead of polygonal templates
and applied their method to similar cell images. All the above methods take
advantage of the Marked Point Process (MPP), in particular, the Area Interaction
Process Prior (AIPP), or any other prior that penalizes the overlapping or
touching, which we explain later in the paper.

Since the structure of the data we are analyzing is different from the
literature, we adapt object representation strategies discussed above to
the problem at hand. When we refer to a shape, we refer to a
family of geometrical objects which share certain features, for
example, an
isosceles and a right triangle both belong to the triangle family. There
are five types of possible shapes of the NPs in our problem. The scientific
reason is that the final shape of the particle is dominated by the
potential energy and the growth kinetics. There is a balance between
surface energy and bulk energy once a nucleus is formed. The
arrangement of
atoms in a crystal determines those energies such that only one of these
specified shapes can be formed. We use similar scientific reasons to
construct shape templates. These templates are determined by the parameters
which vary from shape to shape.

Since there is a difference in the degree of overlapping from
image to image, we assume that the parameters of the AIPP are
unknown and ought to be inferred. This leads to a hierarchical model
setting where the prior distribution has an intractable normalizing
constant. As a result, the posterior is doubly intractable and we use the
Markov chain Monte Carlo (MCMC) framework to carry out the inference.
Simulating from distributions with doubly intractable normalizing constants
has received much attention in the recent literature, but most of these
methods consider the normalizing constant in the likelihood and not in the
hierarchical prior; \citet{Moller06}, \citet{Murray06}, and \citet
{Liang10}, among others. In this paper, we borrow the idea of \citet
{Liang10b}, which is a modified version of the reweighting mixtures
given in
\citet{Chen98} and \citet{Geyer94}, which can deal with doubly intractable
normalizing constants in the hierarchical prior as well. The MCMC algorithm
used can be described as a two-step MCMC algorithm. We first sample the
parameters from the pseudo posterior distribution which is a part of the
posterior that does not contain the AIPP normalizing constant---and
then an
additional Monte Carlo Metropolis--Hastings (MCMH) step that accounts for
this normalizing constant.

Sampling from the pseudo posterior distribution is also quite challenging.
Inferring the unknown number of objects with undetermined shapes is a
complex task. We propose Reversible Jumps MCMC (RJ-MCMC) type of moves to
handle both the tasks [\citet{Green95}]. Birth, death, split and merge
moves have been designed based on the work of \citet{Ripley77a,Rue99}.
We also
propose RJ-MCMC moves to swap (switch) the shape of an object. Using the
above mentioned computational scheme, we obtain the posterior distributions
for all the parameters which characterize the NPs: number, shape, size,
center, rotation, mean intensity, etc. Owing to the model specification
and the computational engine for inferring the model parameters, our
approach extracts the morphological information of NPs, detects NPs laying
on the boundaries, quantities uncertainty in shape classification, and
successfully deals with the object overlapping, when most of the existing
shape analysis methods fail.

The rest of the paper is organized as follows: Section \ref{sec2} describes
the TEM
images, Section \ref{sec3} deals with the object specification procedure,
Section \ref{sec4}
describes the model specification, Section \ref{sec5} describes the MCMC algorithm,
Section \ref{sec6} describes a simulation study and Section \ref{sec7} applies the
method to
the real data. Conclusions are presented in Section \ref{sec8}.

\section{Data}\label{sec2}

\begin{figure}
\begin{tabular}{@{}cc@{}}

\includegraphics{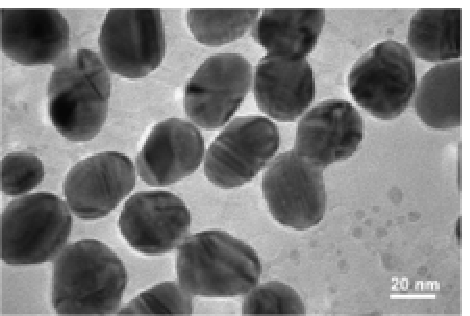}
 & \includegraphics{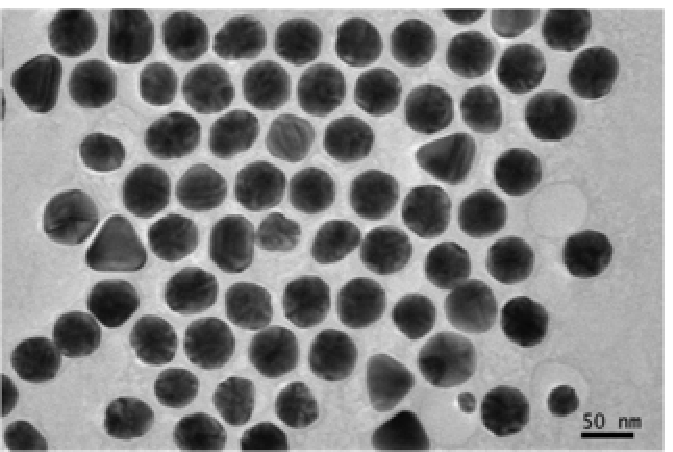}\\
(a) Gold nanoparticles at 20 nm & (b) Gold nano particles at 50 nm
\end{tabular}
\caption{Two examples of TEM nanoparticle images of different
resolution and scale. We observe NPs with different shapes which touch
or slightly overlap each other while many of them lay in the
boundaries.}\label{fig1}
\end{figure}

In this paper we analyze a mixture of gold NPs in a water solution. In
order to
analyze the morphological characteristics, NPs are sampled from this
solution onto
a very thin layer of carbon film. After the water evaporates, the
two-dimensional
morphology of NPs is measured using an Electron microscopy such as TEM.
In our case, a
JEOL $2010$ high resolution TEM operating at $200$ kV accelerating
voltage was used,
which has $0.27$ nm of
point resolution. The TEM shoots a beam of electrons onto the materials embedded
with NPs and captures the electron wave interference by using a
detector on the
other side of the material specimen, resulting in an image. The
electrons cannot
penetrate through the NPs, resulting in a darker area in that part of
the image. The
output from this application will be an eight bit gray scale image
where darker parts
indicate the presence of a nanoparticle. The gray scale intensity is
varying as an
integer between $1$ and $256$. Refer to Figure \ref{fig1} for examples of TEM images.

Due to the absorption of electrons by the gold atoms, the regions
occupied by the
NPs look darker in the image. The darkness pattern may vary according
to specific
arrangements of the atoms inside any single nanoparticle. Additionally, one
can see many tiny dark dots in the background, which are uniformly distributed
throughout the image region. These dark dots are generated because the
carbon atoms
of the carbon film also absorb electrons. One may also notice a white
thin aura
wrapping around the whole or partial boundary of a particle. This is
the result of
having surfactants on the rim of the particles.
The surfactants are added to keep the particles from aggregating in the
process of
making colloidal gold. Analyzing the shapes of the NPs in a TEM image
is primarily
based on modeling them as objects, whose shapes are parametrized.
Treating a
nanoparticle as an object is the critical component of our modeling
framework, which
we discuss in the next section.

\section{Object specification}\label{sec3}
An object is specified in a series of steps that allow us to model a
wide variety of
shapes. They are as follows: (a) template, (b) shift, scale and rotate
operators, (c) object multiplicity. We discuss each of them in detail below.

\subsection{Template}\label{sec3.1}
A template is a predetermined shape which is defined by a set of
parameters which we
call \textit{pure shape parameters} or simply \textit{pure parameters}. We
will call the
template $T$ a \textit{pure object} and we will specify a pure object by
its \textit{pure}
parameters as $g_T^0=\{g_T^0(1),\ldots,g_T^0(q)\}$, where $q$ is the
number of
parameters, and it varies from shape to shape. For example, a circle
with unit
radius at the origin $(0,0)$ can be regarded as a template for circular objects.
Likewise, an equilateral triangle with unit sides, centered at the
origin with the
median aligned to the x-axis, can be a template for triangular objects.
We can
potentially differentiate an equilateral triangle from an isosceles
triangle even
when they both belong to the triangle family. However, to avoid
defining an infinite
number of templates, we consider all types of a particular shape to be
members of
the same template. For example, all types of triangles, such as equilateral,
right-angled, etc., are considered to be members of the triangle
template. As such,
when we refer to a template in this paper we refer to a family of
shapes that has
certain characteristics. A family of shapes is formed by deforming some
of the pure
parameters $\{g_T^0(1),\ldots,g_T^0(q)\}$ in the shape definition. We
distinguish
$g_T^0$ parameters as random (unknown) $g_T^{r}$ and constant (known)
$g_T^{\mathrm{co}}$.
The random pure parameters $g_T^r$ cannot be determined exactly by the
template or
by other components of $g_T^0$. These random pure parameters affect the overall
shape, size and other geometric properties, thereby causing a large scale
deformation of the template. These parameters are closely related to
the template,
but for simplicity we ignore the indicator $T$ and use the notation
$g_T^0=g^0=(g^r,g^{\mathrm{co}})$. The pure parameters are chosen such that the defined
template will have an area equal to the area of a unit circle, that is,
$\pi$ square
units. A template can be shifted, rotated and scaled, still belonging
to the same
shape family.

We also specify landmarks $l^0=l^0(1),\ldots,l^0(M)$ as the $M$ equally spaced
boundary points of a given template. These landmarks can be
determined if one
knows the pure parameters. The landmarks will help us in representing
the shape of the
real image. In polar coordinates, these landmarks can be represented
as
\[
l^0(k)=c_{0,0}+s^0(k)\bigl[\cos\bigl\{\theta(k)
\bigr\},\sin\bigl\{\theta(k)\bigr\}\bigr]^T,
\]
where $s^0(k)$ is
the distance of the $k$th landmark from the center $c_{0,0}$ and
$\theta(k)$ is the
rotation of the $k$th landmark with respect to the baseline. The particular
choice of the coordinate system in which the landmarks are represented
does not
affect~the results. Hence, we have chosen to use polar coordinates for the
simplicity of the mathematical analysis. In this paper we chose ninety landmark
points for all the shapes. Simply speaking, these landmarks in an image
form the
shape. The random deformation of these landmarks results in \textit{small scale}
deformation of the template. In this paper, we focus our attention on
the large
scale deformation since the main goal is to determine the shape and not making
boundary detection or contour tracking, where small-scale deformations are
important. Templates used in the current study are given in (A) in the
online supplementary material [\citet{Konetal13}].

\subsection{Shift, scale and rotation operators}\label{sec3.2}

Apart from the parameters that determine the shape which varies from
template to
template, there are also some common parameters related to shifting,
rotating and
scaling which are needed to represent the actual shape in the image. A
particular
affine shape with shift $c=(c_x,c_y)$, scale $s$ and rotation $\theta$
is given by
the landmarks $l=\{l(1),\ldots, l(M)\}$, whose polar coordinates are
\[
l(k)=c+c_0+s S^0(k)\bigl[\cos\bigl\{\theta(k)+\theta
\bigr\}, \sin\bigl\{\theta(k)+\theta\bigr\}\bigr]^T
\]
for $k=1,\ldots, M$.

\subsection{Object multiplicity and the Markov point process}\label{sec3.3}

In an image we have multiple objects with different shapes and we
assume that the
number of objects is unknown.
A point process is used to model the unknown number of objects and the
overlapping.
One of the widely used models that penalize object overlapping is based
on the
Markov Point Process (MPP) [\citet{Ripley77b}] representation of
objects. In
particular, the Area Interaction Process Prior (AIPP) [\citet
{Baddeley93,Mardia97}] penalized the area of overlap between any two objects.
Below, MPP representation details are given.
The location parameters ${\bc}=(c_{1},\ldots,c_{m})$, the
\textit{points} in the MPP
representation and the number of objects $m$ are modeled as
%
\begin{equation}\label{equ1}
\pi\bigl({\bc},m| \mathbf{g}^r, {\bs},{\bolds{\theta}},{\bT},\gamma
_1,\gamma_2\bigr) =\frac{1}{A^*}\exp\bigl\{-
\gamma_1 m-\gamma_2 S(\bbbeta)\bigr\},
\end{equation}
where $S(\bbbeta)$ denotes the area of the image covered by more than
one object,
$\eta_i=(c_i,s_i,t_i,\theta_i,g_i^r)$ is a collection of parameters
that represents
the $i$th object and $\bbbeta=\bbbeta_{m}=\{\eta_k\}_{k}^{m}$
represents these
parameters for all objects which we call ``object parameters'', $A^*$ is the
normalizing constant which depends on all the parameters described above
$(\bbbeta,m)$ and the positive unknown parameters $\gamma_1$ and
$\gamma_2$
[$A^*=A(\bbbeta,m,\gamma_1,\gamma_2)$]. The interaction parameter
$\gamma_2$
controls the overlapping between objects and $\gamma_1$ the number of
objects in the
image. For example, $\gamma_2=0$ does not penalize overlapping, whereas
$\gamma_2=\infty$ does not allow overlapping at all. Prior
distributions for
$\gamma_1$ and $\gamma_2$ are considered in subsequent sections.
Another way to
penalize object overlapping is the two-way interaction:
\begin{eqnarray*}
\pi({\bc},m| \bbbeta) &=&\frac{1}{A^*}\exp\biggl\{-
\gamma_1 m-\gamma_2 \sum_{i<j}\bigl|R(
\eta_i) \cap R(\eta_j)\bigr|\biggr\}
\\
&&{}\times I[\mbox{no three or more objects have common area}].
\end{eqnarray*}

The indicator term will not allow three or more objects to overlap in
the same area, $R(\eta_i)$ is the region of a
single object characterized by its parameters $\eta_i$ and $R(\eta
_i)\cap R(\eta_j)$ is the overlapping area between the $i$th and the
$j$th object. We
can generalize this case to allow more objects to overlap in a region
and also
penalize with a different parameter $\gamma_k$. Investigating such
models is out of
the scope of this paper. For notational convenience, we introduce
$\bgamma=(\gamma_1,\gamma_2)$ to represent the MPP parameters and
define ${\bT}={\bT_m}\{t_i\}_{i=1}^{m}$, ${\bs}={\bs_m}=\{s_i\}
_{i=1}^{m}$, $\bolds{\theta}=\bolds{\theta}_m=\{\theta_i\}_{i=1}^{m}$,
$\mathbf{g}^r=\mathbf{g}^r_m=\{g_i^r\}_{i=1}^m$ to be used subsequently.

\section{Model}\label{sec4}

\subsection{The likelihood function}\label{sec4.1}

Due to the electron absorption, the mean intensity of the background is
larger than the mean intensity of the regions occupied by the NPs.
Furthermore, since each nanoparticle has different volume size, the
mean pixel intensity for each nanoparticle is different, which is
evident from the representative TEM images of gold NPs shown in
Figure \ref{fig1}. It can also be observed that the overlapping regions usually
have lower intensity because they absorb more electrons in that region.
For tractability, we consider the darkest region to be the dominant
region in determining the configuration of the objects with which it is
overlapping. Due to specific arrangements of the atoms inside any
single nanoparticle, the neighboring pixels have similar intensities.
An appropriate choice for the covariance function in such scenarios is
the popular Conditional Autoregressive (CAR) model [\citet{Cressie93}].
Computationally, a much simpler model is the independent noise model
[\citet{Baddeley93,Mardia97,Rue99}].

After analyzing both real and simulated data sets, the posterior
specification of the
parameters did not change much even if we replaced the CAR model with the
independent Gaussian noise model. An added advantage with the
independent Gaussian
noise model is that it is a lot simpler.
We denote $\bmu=\bmu_{m}=(\mu_{0}, \ldots, \mu_{m})$ as the mean
vector and
$\bsigma^2=\bsigma_{m}^2= (\sigma_{0}^2, \sigma_{1}^2,\ldots,
\sigma_{m}^2)$ as the\vspace*{1pt}
variance vector for the background and objects intensity. To facilitate the
notation, we use $\bTheta=(\bbbeta,m,\bmu,\bsigma^2)$. In this case
the likelihood
can be written as
%
\begin{equation}\label{equ2}
f(Y|\bTheta) \propto\prod_{p=1}^N \exp
\biggl\{-\frac{1}{2\phi(x_p)} \bigl(y_p-\delta(x_p)
\bigr)^2 \biggr\},
\end{equation}
where $N$ is the number of pixels, $x_p$ is the $p$th pixel, $\delta
(x_p)$ is the
mean of the $p$th pixel, $\phi(x_p)$ is the function of the
variance depending on
the pixel and $y_p$ is the intensity of the $p$th pixel. More
explicitly, the mean
intensity for pixels covered by more than one object is taken to be the
minimum mean
intensity of the objects covering the pixels and with variance which
corresponds to
the variance of that object.

For example, in the case where we allow only two-way interaction,
equation (\ref{equ2}) can be
written as
%
\begin{eqnarray}\label{equ3}
&&
f(Y|\bTheta)\nonumber\\
&&\qquad\propto \exp\Biggl\{-\frac{1}{2\sigma_0^2}\sum
_{\nu
\in
R(\eta_0)} (y_{\nu_0}-\mu_{0} )^2-
\sum_{i=1}^m\frac
{1}{2\sigma_i^2}\sum
_{\nu\in
R(\eta_i)\setminus R_{(-i)}} (y_{\nu_i}-\mu_{i} )^2\\
&&\qquad\quad\hspace*{21.3pt}{} -\sum_{i<j}\frac{1}{2\min_{(\mu_{i},\mu_{j})}(\sigma_i^2,\sigma
_j^2)}\sum
_{\nu\in
(R(\eta_i) \cap R(\eta_j))} \bigl(y_{\nu_{i,j}}-\min(\mu_{i},
\mu_{j} ) \bigr)^2 \Biggr\},
\nonumber
\end{eqnarray}
where $R_{(-i)}$ is the region occupied by all objects (NPs) without
the $i$th
object and $R(\eta_0)$ is the area of the background.

\subsection{Prior specification}\label{sec4.2}

We elicit the joint prior distribution hierarchically as follows:
%
\begin{eqnarray}\label{equ4}
\pi(\bTheta,\bgamma)&=& \pi(\bTheta|\bgamma)\pi(\bgamma)
\nonumber
\\
&=& \pi\bigl(\bmu,\bsigma^2\bigr)\pi(\bbbeta,m|\bgamma)\pi(\bgamma)
\\
&=&\pi\bigl(\bmu,\bsigma^2\bigr)\pi\bigl(\bc,m|\bgamma,
\mathbf{g}^r,{\bs},{\bolds{\theta}},{\bT}\bigr)\pi\bigl(\mathbf{g}^r,
{\bs},{\bolds{\theta}},{\bT}\bigr)\pi(\bgamma).
\nonumber
\end{eqnarray}

In the above expression $\pi(\bmu,\bsigma^2)$ is the prior of the
means and the
variances of the background and the objects, $\pi(\bc,m|\bgamma,
\mathbf{g}^r,{\bs},{\bolds{\theta}},{\bT})$ is the joint prior of the
locations and the number of
the objects as given in equation ($1$), $\pi(\mathbf{g}^r, {\bs},{\bolds
{\theta}},{\bT})$ is
the joint prior on all the ``object parameters'' except the locations and
$\pi(\bgamma)$ is the prior on the interaction parameters.

We assume independent $(\mu_i,\sigma_i^2)$ pairs and assign a
noninformative prior
for each of these pairs:
%
\begin{equation}\label{equ5}
\pi\bigl(\bmu,\bsigma^2\bigr)=\prod_{i=0}^m
\pi\bigl(\mu_{i},\sigma_{i}^2\bigr)\propto\prod
_{i=0}^m\bigl(\sigma_i^2
\bigr)^{-1}.
\end{equation}

All the ``object parameters'' except the locations are assumed to be
independent from
object to object. Also, the scale, rotation and template within the
object parameters
are assumed to be independent of other parameters while $g_i^r$ is
assumed to be
closely related to the template $T_i$ (shape). We remind the reader
that $g_i^r$
are different from template to template. In mathematical form we have
%
\begin{equation}\label{equ6}
\pi\bigl(\mathbf{g}^r, {\bs},{\bolds{\theta}},{\bT}\bigr)=\prod
_{i=1}^m \pi(s_i)\pi(
\theta_i)\pi\bigl(g_i^r|T_i
\bigr)\pi(T_i).
\end{equation}

We assign a uniform prior for $s_i$ which is proportional to the size
of the image
$S_{\max}$, that is, $\pi(s_i)\sim U(0,S_{\max})$. All other shapes,
except circles,
have a rotation parameter $\theta\in(0,\pi]$. The prior density for
$\theta$ is
$\pi(\theta) \sim\{|\cos(\theta)|+\pi^{-1}\}/3$, which favors
values near
$\theta=0$ and $\theta=\pi$.
The circle and square do not have a random pure parameter, while the
other considered
templates have at least one random pure parameter. All these parameters
have one
basic characteristic: they are constrained to take values between two variables
$(a_1,a_2)$. We use altered location and scale Beta distribution as a
prior given by
\[
\pi\bigl(g_i^r\bigr)=\frac{1}{\operatorname{Beta}(\alpha,\beta)}
\frac{(g_i^r-a)^{\alpha
-1}(b-g_i^r)^{\beta-1}}{(b-a)^{\alpha+\beta-1}},
\]
where $a,b,\alpha,\beta$ are different for the three different cases.
Furthermore,
we have used the uniform discrete distribution to specify the prior for the
template, $T_i$.

For both the object process parameters $\gamma_1, \gamma_2$ we assume
independent
log-normal distribution priors with parameters which determine a mean
close to $100$
and large variance, $\gamma_1 \sim LN(\alpha_1,\delta_1)$, $\gamma
_2 \sim
LN(\alpha_2,\delta_2)$. We calibrated priors such that inference is as
invariant as possible to changes in the image resolution by defining
parameters in
physical units rather in terms of pixels, and tried to retain their physical
interpretation wherever possible. For example, when we zoom out of an
image, we may
see a great number of objects in the purview, and the perceived.

\subsection{The posterior distribution}\label{sec4.3}
The model proposed above is a hierarchical model of the form
%
\begin{eqnarray}\label{equ7}
y |\bTheta &\sim& f(y|\bTheta),
\nonumber
\\
\bTheta | \bgamma&\sim&\pi(\bTheta|\bgamma) \equiv\frac
{1}{A^*}\pi^*\bigl(
\bc,m|\bgamma,\mathbf{g}^r,{\bs},{\bolds{\theta}},{\bT}\bigr)\pi\bigl(
\mathbf{g}^r,{\bs},{\bolds{\theta}},{\bT}|m\bigr),
\\
\bgamma |\alpha_1,\delta_1,\alpha_2,
\delta_2 &\sim&\pi(\bgamma|\alpha_1,\delta_1,
\alpha_2,\delta_2),
\nonumber
\end{eqnarray}
where $\alpha_1,\delta_1,\alpha_2,\delta_2$ are known values, $A^*$
is a random intractable normalizing constant
and $\pi^*(\bc,m|\bgamma,\mathbf{g}^r,{\bs},{\bolds{\theta}},{\bT})$
is the MPP prior without the
normalizing constant.

The posterior distribution of the parameters
$p(\bbbeta,{\bmu},{\bsigma},m,\bgamma|y)$ is proportional to the
product of (a), (b)
and (c) in the above hierarchical representation:
%
\begin{eqnarray}\label{equ8}
&&
p(\bTheta,\bgamma|y) \nonumber\\
&&\qquad\propto \pi(\bgamma)\pi\bigl({\bmu, \bsigma^2}|
\bbbeta\bigr)\pi(\bbbeta|\bgamma)f\bigl(y|\bbbeta,{\bmu},{\bsigma^2}
\bigr)
\nonumber\\[-8pt]\\[-8pt]
&&\qquad= \frac{1}{A^*}\pi^*\bigl(\bc,m|\bgamma,\mathbf{g}^r,{\bs},{\bolds{
\theta}},{\bT}\bigr)\pi\bigl(\mathbf{g}^r, {\bs},{\bolds{\theta}},{\bT}
\bigr)\pi\bigl(\bmu,\bsigma^2\bigr)\pi(\bgamma)f\bigl(y|\bbbeta,{
\bmu},{\bsigma^2}\bigr)
\nonumber\\
&&\qquad=\frac{1}{A^*}p^*(\bbbeta,{\bmu},{\bsigma},m,\bgamma|y).
\nonumber
\end{eqnarray}
We use the Markov chain Monte Carlo (MCMC) computation algorithm to
carry out
the inference since the posterior distribution is analytically
intractable and the
point process prior has a random intractable normalizing constant. To
facilitate the
discussion, we call $p^*(\bbbeta,{\bmu},{\bsigma},m,\bgamma|y)$ the
pseudo posterior
distribution.

\section{Posterior computation using MCMC}\label{sec5} \label{PostMCMC}

The MCMC algorithm used in this paper can be described as a two-stage
Metropolis--Hastings algorithm. We first sample the parameters from the pseudo
posterior distribution followed by a Monte Carlo Metropolis--Hastings
step to account
for $A^*$ [\citet{Liang10b,LiangBOOK2001}].

The MCMC algorithm will have the following form:
\begin{itemize}
\item Given the current state $\bTheta^k,\bgamma^k$ draw $\bTheta
',\bgamma'$ from
$p^*$ using any standard MCMC sampler.
\item Given all the parameters, simulate auxiliary variables
$z_1,\ldots, z_M$ from the likelihood $z \sim f(z;\bTheta')$ using an
exact sampler.
\item Estimate $R = \frac{A(\bbbeta',m',\bgamma')}{A(\bbbeta^k,m^k,
\bgamma^k)}$ as
\[
\hat{R}=\frac{1}{M}\sum_{i=1}^M
\frac{f(z;\bTheta')}{f(z;\bTheta
^k)}\frac{\pi(\bTheta'|\bgamma')}{\pi(\bTheta^{k}|\bgamma
^{k})}\frac{\pi(\bgamma')}{\pi(\bgamma^{k})},
\]
which is also known as the importance sampling (IS) estimator of $R$.
\item Compute (estimate) the MH rejection ratio $\alpha$ as
%
\begin{equation}\label{equ9}
\hat{\alpha}=\frac{1}{\hat{R}} \frac{p^*(\bTheta',\bgamma
')}{p^*(\bTheta_k,\bgamma_k)}\frac{Q(\bTheta',\bgamma'\rightarrow
\bTheta^{k},\bgamma^{k})}{Q(\bTheta^{k},\bgamma^{k}\rightarrow
\bTheta',\bgamma')}=
\frac{1}{\hat{R}}.
\end{equation}
The last equation is true since $Q=p^*$. So, the above approximates the
normalizing
constant of the posterior.

\item Accept $\bTheta',\bgamma'$ with probability $\min(1;\hat
{\alpha})$.
\end{itemize}

Simulating auxiliary variables $z_i$ from the likelihood is
straightforward and
simply requires us to sample from normal distribution with parameters
defined at the
proposed state of the sampler. The challenge lies in drawing from the pseudo
posterior.

A generalized Metropolis-within-Gibbs sampling with a reversible jump
step is used
to simulate from the pseudo posterior distribution with known number of objects.
Additionally, a reversible jump MCMC (RJ-MCMC) with spatial birth-death
as well as
merge-split move is invoked to sample the number of objects and their
corresponding
parameters.

We draw from the joint pseudo posterior
$p^*(\bmu,\bsigma^2,\bbbeta,\bgamma,m|y)$ by alternately drawing from
the conditional pseudo posteriors of
$\bmu,\bsigma^2\bbbeta|m,y,\bgamma$, $\bgamma|\bmu,\bsigma
^2\bbbeta,m,y$ and $m|\bbbeta,\bmu,\bsigma^2,\bgamma,y$ as follows:

\begin{itemize}
\item Draw $\bbbeta^{k+1},\bmu^{k+1},\bsigma^{k+1}$ from
$p^*(\bbbeta,\bmu,\bsigma|m^k, \bgamma^k, y)$ using a
Metropolis-within-Gibbs
sampler.
\item Draw $m^{k+1}$ from the pseudo posterior
$p^*(m|\bmu^{k+1},\bsigma^{k+1},\bbbeta^{k+1},\bgamma^{k},y)$ using
a RJ-MCMC.
\item Draw $\gamma_{1}^{(k+1)},\gamma_{2}^{(k+1)}$ from the distribution
$p^*(\bgamma|y,\bTheta)$ using an M--H step.
\end{itemize}

We explain these steps in detail, in the following paragraphs.

\subsection{\texorpdfstring{Updating $\eta,\mu,\sigma$, given $m$ and $\gamma$}
{Updating eta, mu, sigma, given m and gamma}}\label{sec5.1}

The conditional distribution of $p^*(\bbbeta|\bmu,\allowbreak\bsigma^2,m,y)$
does not have any
closed form and the same is true for the conditional distribution of
every component
or group of components of $\bbbeta$. A Gibbs sampling step which contains
Metropolis--Hastings steps and RJ-MCMC steps is utilized. In the online
supplementary material (B) [\citet{Konetal13}] we give the
Metropolis--Hastings updates for $(\bbbeta,\bmu,\bsigma)$ excluding
$\mathbf{T}$, which
is given next.

\subsubsection{Updating the template $T_j$ (swap move)}\label{sec5.1.1}

We can view the problem of shape selection as a problem of model
selection between
$M_{j,t_1},\ldots,M_{j,t_D}$, where $M_{j, t_i}$ represents the model
with template
$t_i$. Moving from shape to shape is considered a difficult task since
not only the
pure parameters that characterize the template are different, but also
the parameter
specification may not have the same meaning across templates. For
example, one can
argue that the scaling parameter of a circle can be different from the
scaling parameter of a triangle. The move from shape to shape is based
on the rule
that both shapes should have the same area and the centers of both
shapes are the
same. This increases the likelihood of generating good proposals. For
the particular
shapes we deal with, the equality of area also means equality of the scaling
parameter. This means that all of the above models $M_{j,t_i}$ have the
same scaling
$s_j$ and location $c_j$ parameters. The rotation parameter, $\theta$,
can be chosen
such that the proposed shape overlap ``matches'' as much as possible to
the existing
shape given the same $(s_j, c_j)$ or simply one may retain the same
$\theta$ while
changing shapes. The \textit{pure random} parameters are the only
parameters that do not
have a physical meaning when we change the shape and also their number
could vary
from shape to shape. Reversible Jump MCMC is used successfully for
problems with
different dimensionality and is characterized by introducing auxiliary
variables for
the unmatched parameters [\citet{Green95}]. This is the approach we
follow in this paper. For more details see Appendix \ref{appA}.

\subsection{Updating $m$}\label{sec5.2}
Two different types of moves are considered in updating the number of objects:
birth-death and split-merge. In the death step, one chosen-at-random
object is
deleted and in the birth step, one object with parameters generated
from the priors
is added. In the merge step we consider the case where two objects die
and give
birth to a new one and in the split step two new objects are created in
the place of
one. For more detail see Appendix \ref{appB}.

\subsection{\texorpdfstring{Updating $\gamma$}{Updating gamma}}\label{sec5.3}

The random walk $\log$-normal proposal 
is used to sample
from the pseudo posterior distribution of $\bgamma$, $p^*(\bgamma
|\bTheta,y)$.

\section{Simulations}\label{sec6}
\begin{figure}
\begin{tabular}{@{}cc@{}}

\includegraphics{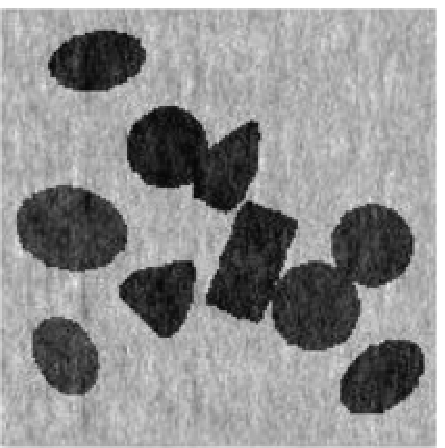}
 & \includegraphics{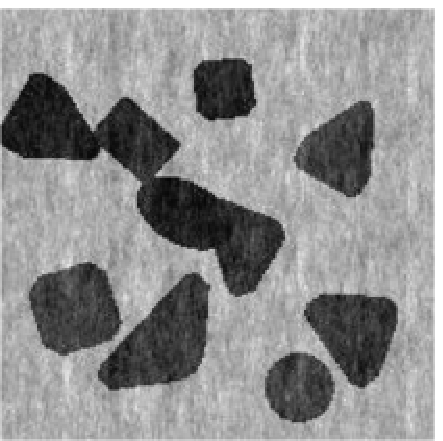}\\
(a) $\gamma_2=40$ & (b) $\gamma_2=10$
\end{tabular}
\caption{Two different simulated images with ten objects, $m=10$, and
two different
values for the interaction parameter $\gamma_2$. The value of the
interaction parameter is related to the degree of overlapping.}
\label{fig2}
\end{figure}

\begin{figure}[b]
\begin{tabular}{@{}c@{\hspace*{6pt}}c@{}}

\includegraphics{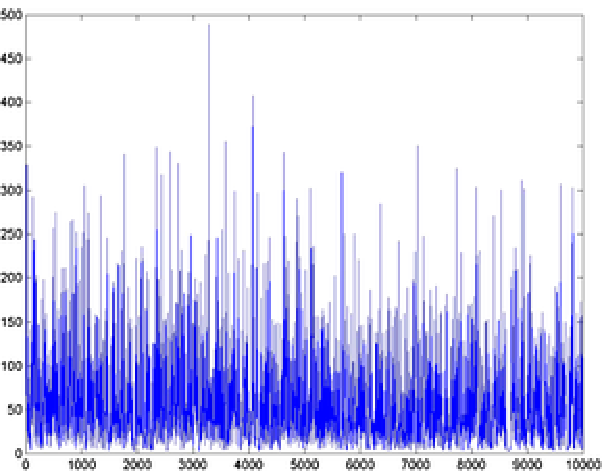}
 & \includegraphics{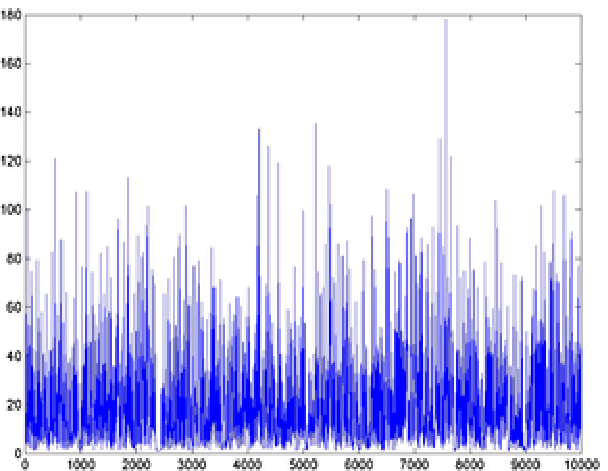}\\
(a) $\gamma_2$ posterior & (b) $\gamma_2$ posterior
\end{tabular}
\caption{Trace plot of the last $10^4$ MCMC sample values from the
posterior of $\gamma_2$ for the two different simulated images:
\textup{(a)} for the first image where $\gamma_2=40$ and \textup{(b)} for the second
image where $\gamma_2=10$. The MCMC for both cases converges to right
skewed distributions with different medians.}
\label{fig3}
\end{figure}

In this section we use a simulation study to evaluate the performance
of our
proposed MCMC method. We consider two examples wherein a $200 \times
200$ image
with ten objects each are generated from the prior distributions described
in Section \ref{sec4.2} with area interaction parameter $\gamma_2=40$ and
$\gamma_2=10$, respectively. The pixels inside each object have
constant mean, which
is different from object to object. The covariance matrix is chosen
from a CAR model
with parameters very close to the extreme dependence. Objects in both
the example
images have different morphological properties and belong to the five different
shape families described in Section \ref{sec2}. The image used in the first
example is
shown in Figure \ref{fig2}(a) and the second is shown in Figure \ref{fig2}(b). For the
%
example images,
the MCMC samples drawn from the posterior distribution of $\gamma_2$
are given in
Figure \ref{fig3}. From these simulations, we can see that the Markov chain
mixes well and the
posterior mean is close to the true values we used to simulate the
data. Values
close to $40$ are drawn in example 1 [Figure \ref{fig1}(a)], while values close
to $10$ are
drawn in example 2 [Figure \ref{fig2}(b)]. A general observation in the
%
simulations is that the
variance of the posterior distribution of $\gamma_2$ depends on the
value of
$\gamma_2$. For large values of $\gamma_2$ we observe relatively
larger posterior
variance than for small values. Another significant observation is that there
is a dependence on the accuracy and the variance of the posterior
distribution of
$\gamma_2$ on the number and size of objects. To investigate this
phenomenon, we
fixed the value of $\gamma_2$ but simulated images with a different
number of objects
and sizes. As we increase the number and the size of objects, the posterior
distribution of $\gamma$ will be closer to the true value. Below, we
discuss two
features of our method using the two examples.

\subsection{Unknown AIPP parameters}\label{sec6.1}
\begin{figure}
\begin{tabular}{@{}cc@{}}

\includegraphics{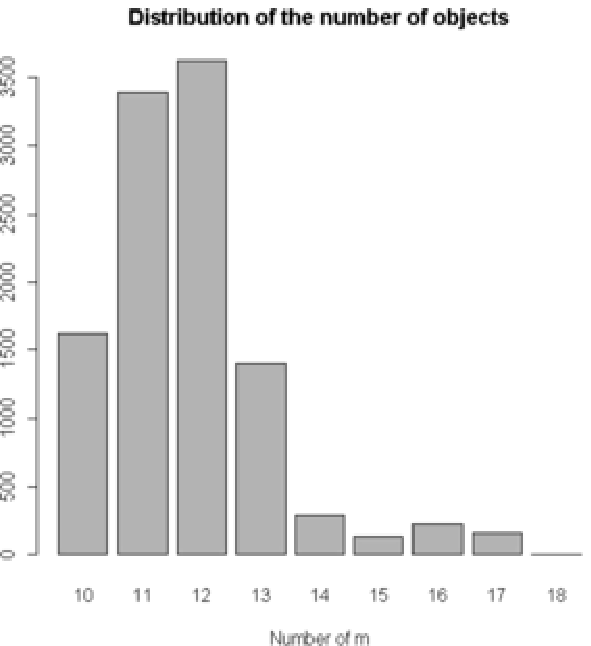}
 & \includegraphics{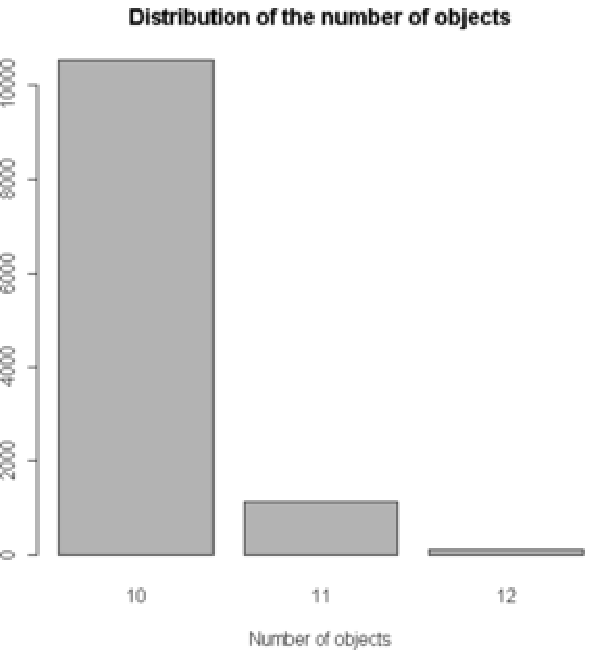}\\
(a) $\gamma_2=0$ & (b) $\gamma_2$ random
\end{tabular}
\caption{Distribution of the number of objects, $m$, considering \textup{(a)}
$\gamma_2=0$
and \textup{(b)} $\gamma_2$ random. Considering an area interaction penalty in
our application will improve the convergence of the MCMC algorithm to
the right number of objects.}
\label{fig4}
\end{figure}

We demonstrate one of the advantages of treating the AIPP parameters as unknown.
First, we compare the MCMC results from the proposed model with the
results of the
model that does not penalize overlapping. More specifically, we treat
$\gamma_2$ as
a random variable in the first scenario, and then consider it known and
misspecified
in the second scenario. In both the runs, the parameter $\gamma_1$ is
set to its
true value $10$. The MCMC posterior distribution of $m$ for the image
in Figure \ref{fig2}(a), in a
total of $12\mbox{,}000$ iterations, is recorded and presented for these two
different cases
in Figure \ref{fig4}. The distribution of the number of objects $m$ in the case
%
of $\gamma_2=0$
is mostly a misspecification of the real image. In this case we have a
sample of up
to $18$ objects, which is almost twice the original number of objects.
An obvious
overestimation of the number of objects in the posterior distribution
occurs when we
do not penalize the overlapping. On the other hand, when we choose
$\gamma_2$ as a
random variable $90\%$ of the posterior simulated number of objects
represent the
true number of objects. Treating $\gamma_2$ as unknown, in comparison with
$\gamma_2=0$, yields a better fit and improves classification. For the
case where
$\gamma_2$ is fixed at a value different from zero, the answer depends
on how close
the original and the assumed value of $\gamma_2$ are. If we fix the
value of
$\gamma_2$ in the range determined from the MCMC updates, the results
on the number
of particles and shape analysis are not very different from the
original values.
Nevertheless, values outside the range can change the results
dramatically. The same
observations are true for the second simulated image [Figure \ref{fig2}(b)] as well.

\subsection{Split and merge moves}\label{sec6.2}
\begin{figure}
\begin{tabular}{@{}cc@{}}

\includegraphics{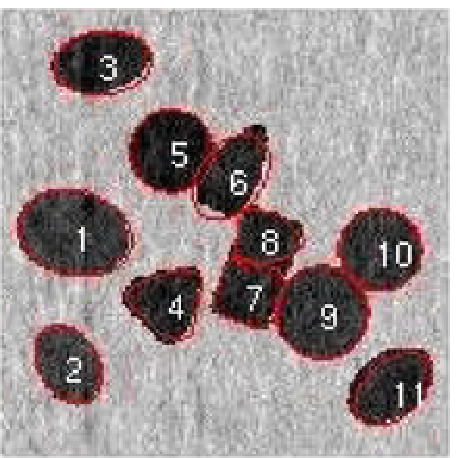}
 & \includegraphics{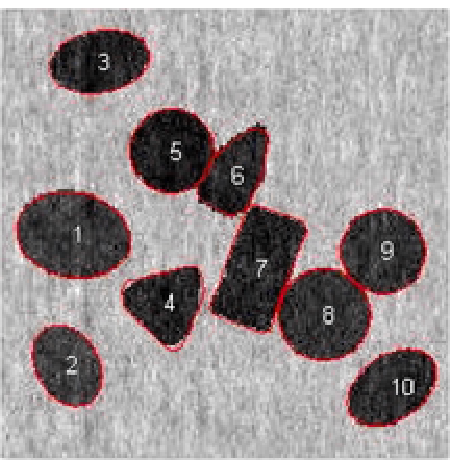}\\
(a) $1000$th iteration & (b) $1500$th iteration
\end{tabular}
\caption{Merge step in action: object configuration \textup{(a)} before merge
\textup{(b)} after merge. We chose MCMC movements with low acceptance
probability ratio to show the success of our method. There is always a
chance for the algorithm to be trapped into local minima if we do not
use the right MCMC moves and proposals.}
\label{fig5}
\end{figure}

\begin{figure}[b]
\begin{tabular}{@{}cc@{}}

\includegraphics{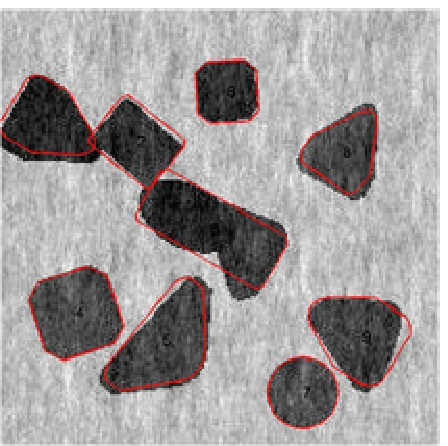}
 & \includegraphics{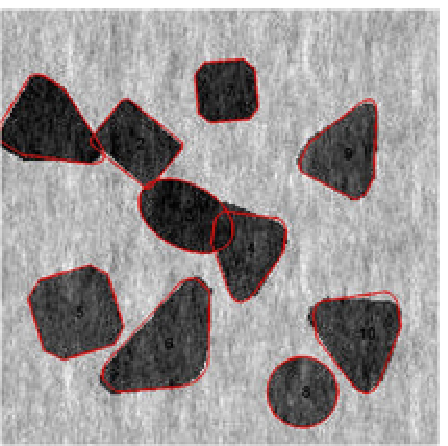}\\
(a) $1400$th iteration & (b) $1700$th iteration
\end{tabular}
\caption{Split step in action: object configuration \textup{(a)} before split
\textup{(b)} after split. It is obvious that other MCMC movements have occurred
as well since shape parameters such as location, size and rotation have
changed.}
\label{fig6}
\end{figure}

Another feature of our proposed method is the split and merge type of
move. We can
see the merge and split step in action in Figures \ref{fig5} and \ref{fig6},
respectively. In the
absence of this type of move, it would have required a large number of MCMC
iterations to arrive at the configurations shown. We present the two
different move
%
steps that occurred in the two simulated images. The $1000$th and the
$1500$th MCMC
iteration is given for the first image. In addition to different
changes that have
occured, there is an obvious merge move step, wherein the seventh and eighth
objects in Figure \ref{fig5}(a) are merged to form the seventh object in
%
Figure~\ref{fig5}(b). Similarly,
we show the split move in action using example 2. Snapshots taken at the
$1400$th and the $1700$th MCMC iterations for example 2 are given in
Figure \ref{fig6}(a) and (b). Not only an obvious split step has occurred but also we
can see the
different deviations of the boundaries which are related to the object
representation parameters.

\subsection{Implementation details}\label{sec6.3}

All the simulations and the algorithms were implemented in MATLAB,
running on a Xeon
dual core processor clocking $2.8$ GHz with $12$ GB RAM. MCMC chains are
initialized
by using classical image processing tools. All the five templates are randomly
assigned to complete template specification. The simulation time for
the two
examples is approximately two hours for $12\mbox{,}000$ iterations. Convergence
of the
chains was observed within the first $1000$ iterations. However, we
point out that
the computational time of the proposed method depends on the size of
the image, the
number of the objects and the complexity of overlapping, and burn-in
time which
strongly depends on the initial state of the chain.

In order to accelerate quick mixing, we take advantage of several
classical image processing tools. Notable among
these are the watershed image segmentation and certain morphological
operator based image
filtering techniques such as erosion, dilation etc. [\citet
{Gonzalez08}]. For example,
we use watershed segmentation to decompose the image into subimages
that have
approximately nonoverlapping regions (in terms of objects). A repeated
application
of the erosion operator on the subimages, in conjunction with
connected-component
analysis and dilation operation, gives us an approximate count of
number of objects and
their morphological aspects. Such information can be used to initialize
the chains
and to construct proposal distributions required by the MCMC sampler.
In addition,
the region-based approach allows one to exploit distributed and
parallel computing
concepts to reduce simulation time and make the algorithm scalable.
Further details
are not presented here since morphological preprocessing is not the
subject of the
present work. We point out above that choices affect simulation time
and may improve mixing
but otherwise are not necessary for our proposed method to work. In addition,
simulation time and effort required by the MCMC method required are
relatively small
compared to the time, effort and resources required to produce the NPs
and finally
obtain the TEM images which can exceed weeks.

\section{Application to gold nano particles}\label{sec7}

Using the MCMC samples, we can obtain the distribution of the particle
size, which
is characterized by the area of the nanoparticle and the distribution
of the
particle shape. The aspect ratio, defined as the length of the
perimeter of a
boundary divided by the area of the same boundary, can be derived from the
combination of size, shape and the pure parameters. The statistics of
size, shape
and aspect ratio are widely adopted in nano science and engineering to
characterize
the morphology of NPs, and are believed to strongly affect the physical
or chemical
properties of the NPs [\citet{El01,Nyiro09}]. For example, the
aspect ratio is considered as an important parameter relevant to
certain macro-level
material properties because physical and chemical reactions are
believed to
frequently occur on the surface of molecules so that as the aspect
ratio of a
nanoparticle gets larger, those reactions are more active.

We apply our method to three different TEM images. The parameters that
maximize the
posterior distribution (MAP) obtained from the (MCMC) are presented in
detail. Our
classification results of particular type are verified by our
collaborators with
domain expertise; this manual verification appears the only valid way
for the time
being. More than $95\%$ of the NPs in those images are classified
correctly. This
also includes the particles in the boundary as well as having
overlapping regions.
For completely observed objects, there is almost $100\%$ correct classification.

We start our application with the image in Figure \ref{fig1}(a). Morphological
image processing operations, such as watershed transformation and
erosion, can be used to get an approximate count of the number of NPs
in the model [\citet{Gonzalez08}]. They also can be used in
initializing the MCMC chains and in constructing proposal distributions
required by the MCMC sampler. The morphological image processing we
used in this dissertation has the following steps: (1) image filtering
and segmentation, (2) determining the number of objects, (3)
estimating location, size and rotation parameters. We first transform
the image from grey to a binary image and then apply watershed
transformation to partition the image into subimages. In each binary
subimage we apply erosion and dilation operations to find initial
values for the parameters inside of each subimage. Because this
morphological processing is not the subject of the present work, it is
not presented in more detail. After the initial values are obtained
from the preprocessing step, all five templates are randomly assigned
for starting template specifications. From the MCMC sampler described
in Section \ref{PostMCMC} we obtain a random sample of the posterior
distribution for all the parameters which characterize the NPs, namely,
the \textit{shape} $\bT$, the \textit{size} $\bs$, the
\textit{rotation} $\bolds{\theta}$, the \textit{random pure parameter}
$\mathbf g^r$, the \textit{mean} intensity $\bmu$ and the \textit{variance}
$\bsigma^2$. We use this posterior sample for inferring the model
parameters and extracting the morphological information of NPs with
uncertainty in shape size and classification. To better present our
results, we chose to work with the Maximum a posteriori (MAP)
estimations of these parameters.

\begin{figure}

\includegraphics{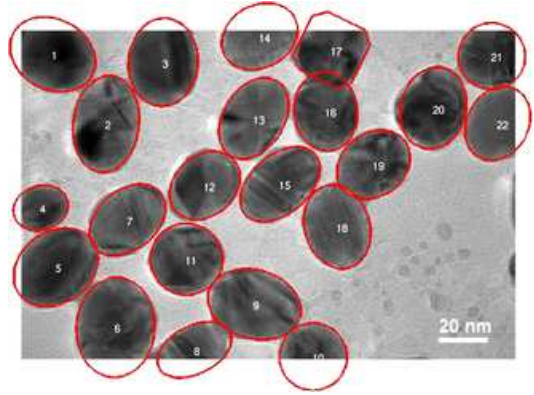}

\caption{Example 1: maximum a posteriori estimation
using 20,000 MCMC samples. The proposed method can deal successfully
with overlapping and boundary objects since $11$ out of $22$
nanoparticles in the image are in the boundaries.}
\label{fig7}
\end{figure}

\begin{figure}
\begin{tabular}{@{}cc@{}}

\includegraphics{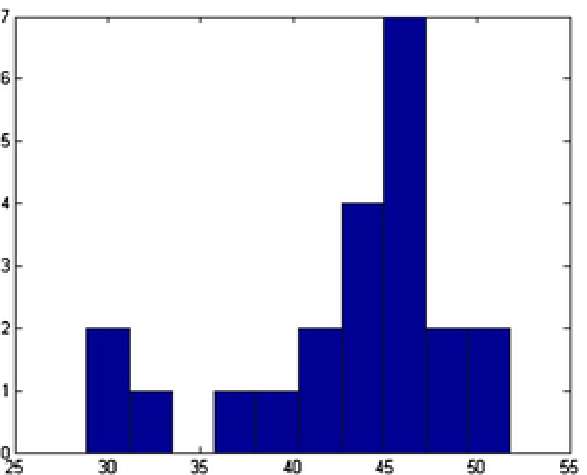}
 & \includegraphics{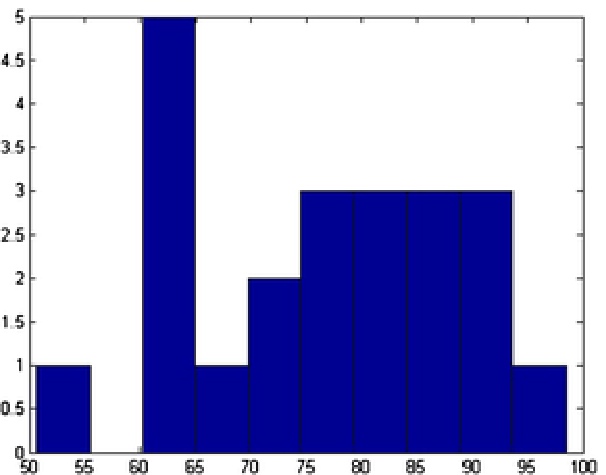}\\
(a) $s$ (scale) & (b) $\mu$ (foreground intensity)\\[6pt]
\multicolumn{2}{@{}c@{}}{
\includegraphics{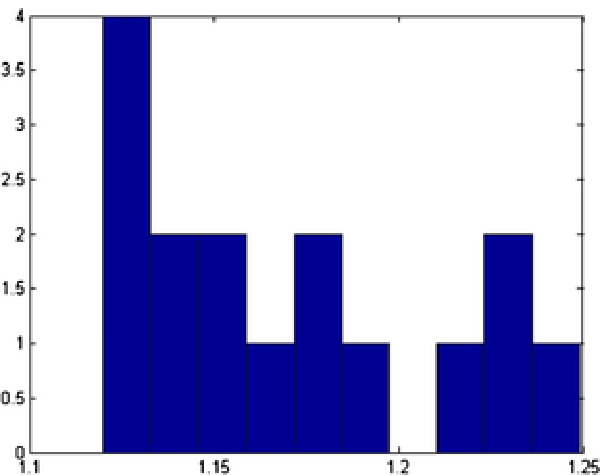}
} \\
\multicolumn{2}{@{}c@{}}{(c) $g^{r}$ (random pure parameter)}
\end{tabular}
\caption{Distribution of the MAP estimates for shape parameters in
example 1. Specifically: \textup{(a)}
the scale, \textup{(b)} the mean intensity for different objects and \textup{(c)} the
pure parameters
for ellipse.}
\label{fig8}
\end{figure}

In Figure \ref{fig7} we show the TEM image and MAP estimates of the parameters
for $20\mbox{,}000$ MCMC
sample. In Figure \ref{fig8} we present the parameters of $\bs$, $\bg^r$ and
$\bmu$ that correspond to the
MAP estimate for all the number of objects, $m$, corresponding to that value.
Summary statistics of the shape parameters are given in Table \ref{tabl1}.
From the table and the histogram it is clear that the mean intensity is
different
from nanoparticle to nanoparticle, justifying our assumption of
different means in~(\ref{equ3}). We also obtain the posterior probability of the
classification for each of
the objects. This probability depends on the complexity of the shape of
the object.
For example, object $2$ has been classified as an ellipse with
probability $0.98$,
whereas object $20$ has been classified as an ellipse with probability
$0.68$ (circle
with probability~$0.32$). In Table \ref{tabl1} (and in all the following tables
of this
chapter) we presented the classification with the highest posterior
probability of some of the nanopartiles. In this example we
successfully deal with the object overlapping and objects laying
on the boundaries.

%
\begin{table}
\caption{MAP estimates of the parameters for the first six objects in
example 1}\label{tabl1} 
%
\begin{tabular*}{\tablewidth}{@{\extracolsep{\fill}}l c c c d{2.2} c c@{}} 
\hline
\textbf{Object} & \textbf{Shape} $\bolds{(T)}$
& \textbf{Center} $\bolds{(x,y)}$ & \textbf{Size} $\bolds{(s)}$
& \multicolumn{1}{c}{\textbf{Rotation} $\bolds{(\theta)}$} &
$\bolds{g^{r}}$ & \textbf{Mean}
$\bolds{(\mu)}$ \\  
\hline
1 & E & $(39.68, 32.72)$ & 51.49 & -0.21 & 1.14 & 50.64\\
2 & E & $(105.92, 105.92)$ & 49.41 & 1.41 & 1.22 & 74.67 \\
3 & E & $( 175.79, 41.29)$ & 47.20 & 1.36 & 1.12 & 62.55 \\
4 & E & $(25.87, 221.72)$ & 28.86 & 0.61 & 1.15 & 71.58\\
5 & E & $(39.89, 297.00)$ & 49.98 & 0.83 & 1.13 & 64.58\\
6 & C & $(116.07, 362.30)$ & 51.82 & \multicolumn{1}{c}{NA} & \multicolumn{1}{c}{NA} & 73.76\\
\hline
\end{tabular*}
%
\end{table}

\begin{figure}[b]

\includegraphics{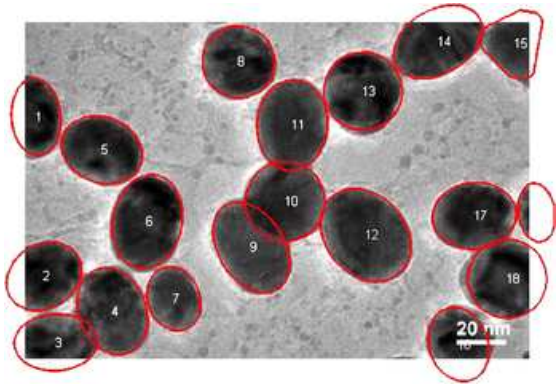}

\caption{Example 2: maximum a posteriori estimation
using 20,000 MCMC samples. Our method has distinguished the
nanoparticles even in the case when they overlap in groups such as
9--10--11--12 nanoparticles.}
\label{fig9}
\end{figure}

\begin{figure}[b]
\begin{tabular}{@{}cc@{}}

\includegraphics{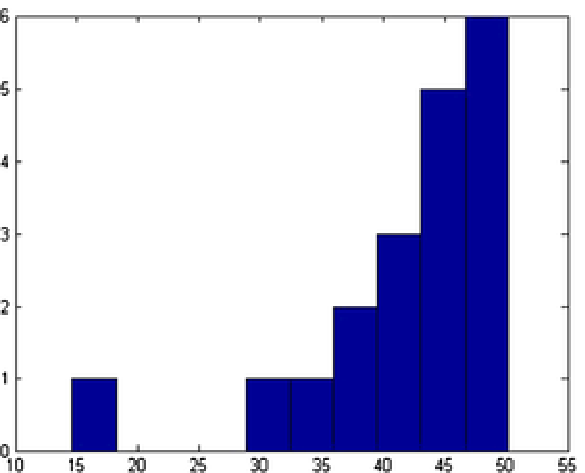}
 & \includegraphics{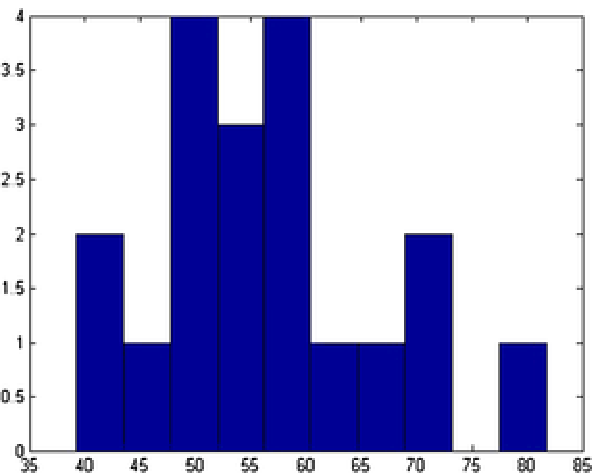}\\
(a) $s$ (scale) & (b) $\mu$ (foreground intensity)\\[6pt]
\multicolumn{2}{@{}c@{}}{
\includegraphics{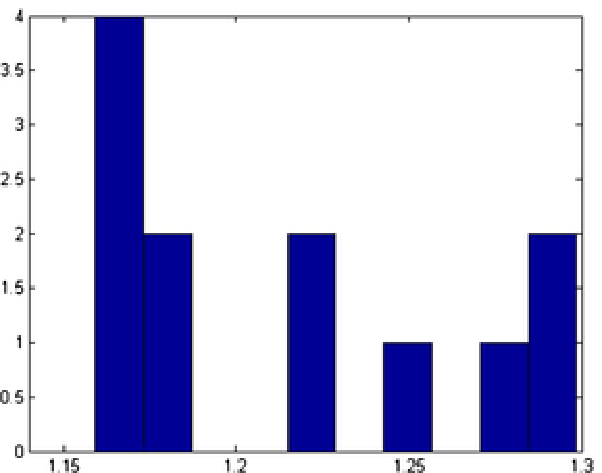}
} \\
\multicolumn{2}{@{}c@{}}{(c) $g^{r}$ (random pure parameter)}
\end{tabular}
\caption{Distribution of the MAP estimates for nanoparticle parameters
in example 2.
Specifically: \textup{(a)} the scale, \textup{(b)} the mean intensity for different
objects and \textup{(c)} the pure
parameters for ellipse.}
\label{fig10}
\end{figure}

Our second application deals with a more complex image shown in
Figure~\ref{fig9}. In this
image at least $6$ overlapping areas and at least $6$ nanoparticles
laying in the
boundary are observed. More specifically, nanoparticles
$1$, $2$, $3$, $14$, $15$, $16$, $18$, and $19$ lay in the boundary of the
image while
pairs 2--4, 3--4, 9--10, 10--11, 17--18, and 10--12 overlap. In this
example, the
overlapping is more complex and existing methods fail to represent the real
situation. A~number of nanoparticles are overlapped together forming a
groups such as nanoparticles 9--10--11--12. MAP estimate values for all
the parameters are obtained after $20\mbox{,}000$ MCMC
iterations. Complex shapes have been classified accurately; see
Figure \ref{fig9}. For example,
nanoparticle $18$ has an incomplete image and it has been classified as
a circle
with posterior probability $0.77$. The MAP estimates of the parameters
drawn from
MCMC, namely, shape $\bT$, size $\bs$, rotation $\bolds{\theta}$,
random pure parameter $\bg^r$ and
mean intensity $\bmu$ are presented for the first six objects in
Table \ref{tabl2}. In this
application, $11$~out of the $17$ objects are ellipses (E) and $6$ are
circles (C)
and one is a triangle~(TR). We also present the histogram of the MAP
estimates of
parameters $\bs$, $\bg^r$ and $\bmu$ in Figure \ref{fig10}. Summary
statistics of various shape
parameters are given in Table~\ref{tabl2}. We see from the table that our
proposed algorithm
captures triangles, circles etc. quite accurately.

\begin{table}
\caption{MAP estimates of the parameters for the first six objects in
example 2}\label{tabl2} 
%
\begin{tabular*}{\tablewidth}{@{\extracolsep{\fill}}l c c c d{2.2} c c@{}} 
\hline
\textbf{Object} & \textbf{Shape} $\bolds{(T)}$
& \textbf{Center} $\bolds{(x,y)}$ & \textbf{Size} $\bolds{(s)}$
& \multicolumn{1}{c}{\textbf{Rotation} $\bolds{(\theta)}$} &
$\bolds{g^{r}}$ & \textbf{Mean}
$\bolds{(\mu)}$ \\  
\hline
1 & E & $(13.97, 256.78)$ & 37.48 & -1.51 & 1.2960 & 39.185\\
2 & C & $(27.44, 275.96)$ & 41.04 & \multicolumn{1}{c}{NA} & NA & 42.969 \\
3 & E & $(37.56, 314.44)$ & 38.02 & -0.29 & 1.2175 & 52.569 \\
4 & E & $(106.40, 321.61)$ & 47.44 & -1.17 & 1.1591 & 60.605\\
5 & E & $(93.20, 413.87)$ & 44.33 & -0.36 & 1.1612 & 51.080\\
6 & E & $(146.67, 406.42)$ & 49.63 & -1.76 & 1.1621 & 44.617\\
\hline
\end{tabular*}
%
\end{table}
%

Our next application deals with an image with $76$ nanoparticles with
$4$ shapes;
see Figure \ref{fig1}(b).
In this image, few objects have overlapping areas and at least $10$
nanoparticles
are laying in the boundary.
Some objects do not have very clear shape like objects $29$ and $31$.

\begin{figure}

\includegraphics{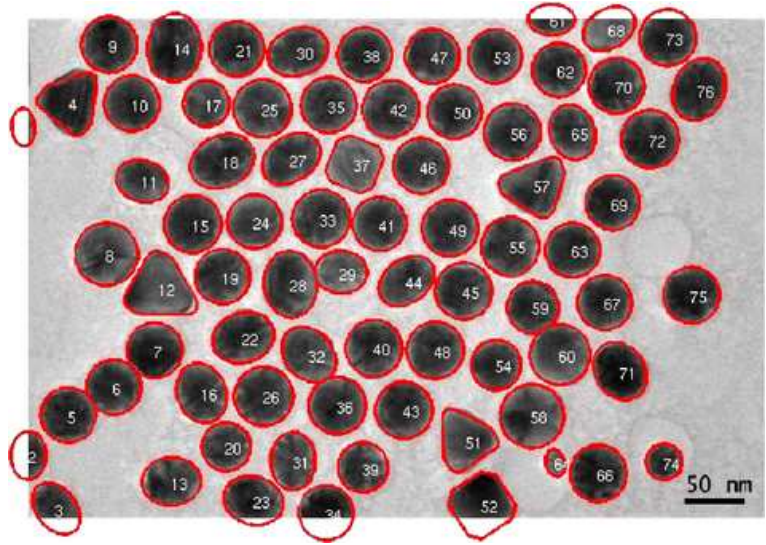}

\caption{Example 3: maximum a posteriori estimation using 20,000 MCMC
samples. The proposed method has successfully classified the shape of
all nanoparticles in the image counting also for uncertainty.}
\label{fig11}
\end{figure}

\begin{figure}
\begin{tabular}{@{}cc@{}}

\includegraphics{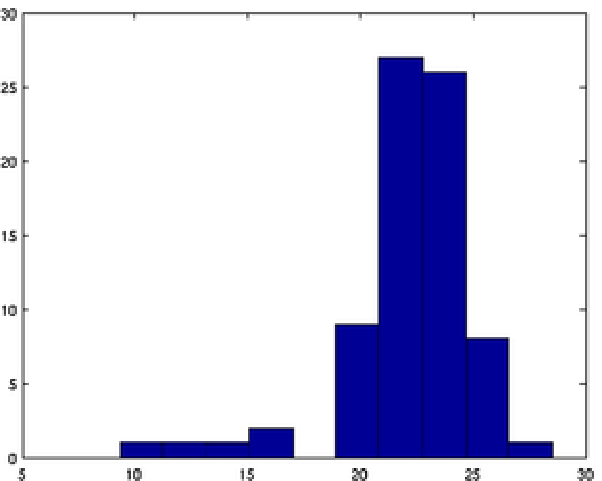}
 & \includegraphics{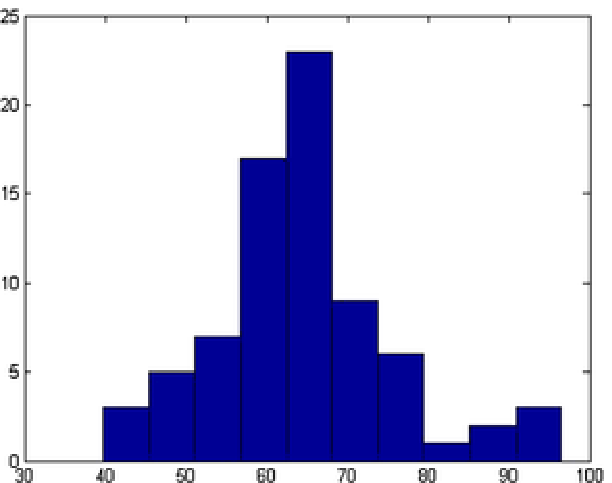}\\
(a) $s$ (scale) & (b) $\mu$ (foreground intensity)\\[4pt]
\multicolumn{2}{@{}c@{}}{
\includegraphics{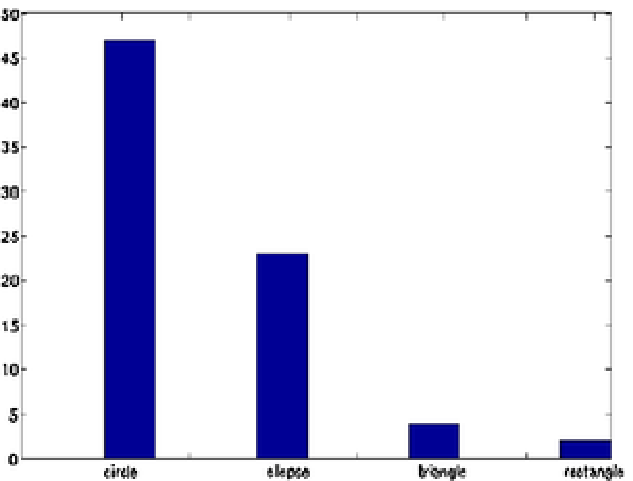}
} \\
\multicolumn{2}{@{}c@{}}{(c) $T$ (shape classification)}\vspace*{-3pt}
\end{tabular}
\caption{Distribution of the MAP estimates for shape parameters in
example 3. Namely, the
distribution of: \textup{(a)} the scale, \textup{(b)} the mean intensity for different
objects and \textup{(c)}
the shape classification.}
\label{fig12}\vspace*{-3pt}
\end{figure}

Different shapes are captured with different templates with the
proposed method. In addition to the circles and ellipses which were
successfully captured in the previous images, the triangles and squares
are also captured accurately. Nanoparticles denoted by $29$ and $31$
are classified correctly, even if they have vague shapes; see Figure
\ref{fig11}. In this example, out of $76$ nanoparticles, $47$ are
classified as a circle, $23$ as an ellipse, $4$ as a triangle and $2$
as a square. Distribution of the various parameters of the identified
objects are shown in Figure \ref{fig12}. In Table~\ref{tabl3} we
present all the triangular shapes in order to compare the pure
parameter $h_1$. As we can see from the table, triangular shape
nanoparticles $4$ and $12$ are closer to the equilateral triangle, with
value close to $h_1=2.33$, while triangular shape nanoparticles $51$
and $57$ have wider sides, since their $h_1<2.3$.

\begin{table}[b]\vspace*{-3pt}
\def\arraystretch{0.9}
\caption{MAP estimates of the parameters for the first six objects in
example 3}\label{tabl3} 
%
\begin{tabular*}{\tablewidth}{@{\extracolsep{\fill}}l c c c d{2.2} c c@{}} 
\hline
\textbf{Object} & \textbf{Shape} $\bolds{(T)}$
& \textbf{Center} $\bolds{(x,y)}$ & \textbf{Size} $\bolds{(s)}$
& \multicolumn{1}{c}{\textbf{Rotation} $\bolds{(\theta)}$} &
$\bolds{g^{r}}$ & \textbf{Mean}
$\bolds{(\mu)}$ \\  
\hline
\hphantom{0}1 & E & $(-3.11, 68.18)$ & 12.43 & -1.57 & 1.29 & 66.27\\ 
\hphantom{0}4 & $T$ & $(35.53, 110.92)$ & 25.82 & 1.38 & 2.32 & 49.33\\
12 & $T$ & $(306.90, 225.73)$ & 28.73 & 0.35 & 2.31 & 79.59 \\
28 & E & $(219.91, 221.35)$ & 24.09 & 1.53 & 1.14 & 68.19\\
51 & $T$ & $(365.75, 352.49)$ & 24.61 & -1.46 & 2.25 & 63.29\\
57 & $T$ & $(422.15, 139.28)$ & 25.25 & 0.25 & 2.01 & 70.49\\
\hline
\end{tabular*}
%
\end{table}

In this image we can see more than $85\%$ percent of the nanoparticles
are in the same
shapes like circular or slightly tilted like ovals. Normally when we do shape
controlled synthesis, we called it nano spheres or circular nanoparticles.
Approximately five to ten percent of the other shapes or slight changes
we usually neglect
because in solution synthesis routes it is very difficult to synthesis
$100\%$ of the same
size and same shapes. However, if we consider critically the reason of shape
evolution or statistical analysis of different shapes, then this small
difference
might be considered. We classify this particular example as spherical gold
nanoparticles having almost the same size and shapes.

\begin{figure}

\includegraphics{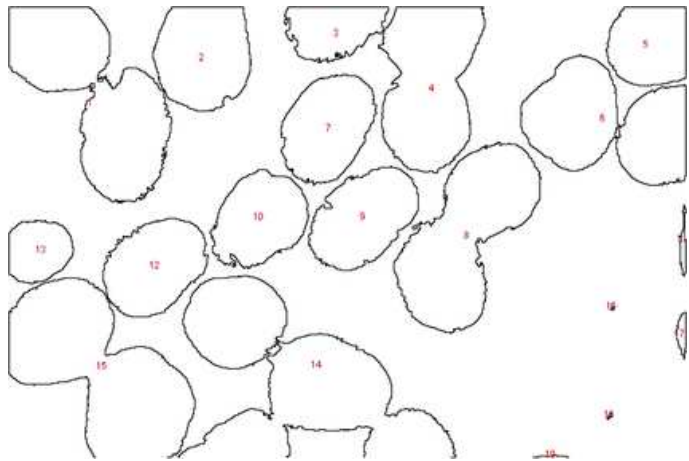}

\caption{Objects identified by ImageJ in example 1. Out of the $22$
particles, $4$ are
recognized. Recognition $\mbox{rate} = 18.18$\%.}
\label{fig13}
\end{figure}

\begin{figure}[b]

\includegraphics{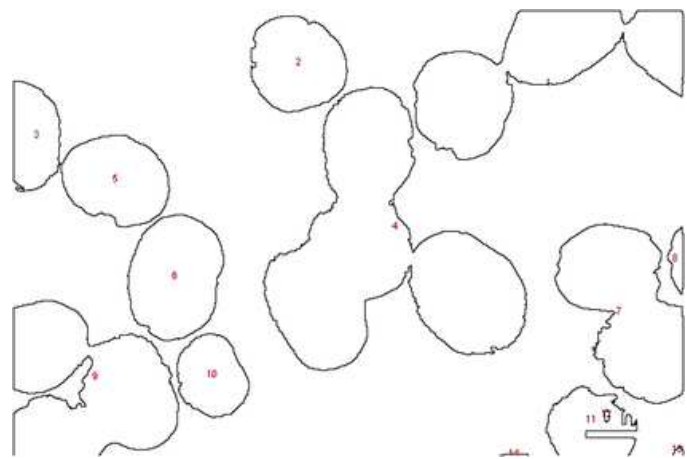}

\caption{Objects identified by ImageJ in example 2. Out of the $19$
particles, $6$ are
recognized. Recognition $\mbox{rate} = 35.58$\%.}
\label{fig14}
\end{figure}

As a part of the verification process, we compare the accuracy of our
method with
that of the current practice used in nanoscience.
In brief, the current practice is largely a manual process with support
of image
processing tools
such as ImageJ Particle Analyzer (\url{http://rsbweb.nih.gov/ij}) and AxioVision
(\url{http://www.zeiss.com/}),
which have been popularly used for biomedical image processing. The
results are
shown in Figures \ref{fig13} and \ref{fig14}.

The manual counting process, subject to the application of the above
imaging tools,
is necessitated by the low accuracy of the autonomous procedures. For
three TEM images with overlaps
among particles, our procedure recognized $95\%$ of the
total articles compared to the 20--50\% recognition rate of the ImageJ.
Considering frequent occurrence of overlaps in the TEM images of
nanoparticles, the
existing software cannot be used as more than a supporting tool. We
have also applied our method to other images with the same success,
encouraging its applicability.

\section{Conclusion}\label{sec8}
We adopted a Bayesian approach to image classification and segmentation
simultaneously and applied it in TEM images of gold nanoparticles. The
merit of our
development is to provide a tool for nanotechnology practitioners to
recognize the
majority of the nanoparticles in a TEM image so that the morphology
analysis can be
performed subsequently. This can evaluate how well the synthesis
process of
nanoparticles is controlled, and may even be used to explain or design certain
material properties. Several factors like kinetic and thermodynamic
parameters, flux
of growing material, structure of the support, presence of defects and
impurities can
affect the morphology of NPs. In the future, we are planning to perform
a factorial type
experiment to identify the significant factors for morphological study. These
significant factors can be properly controlled to develop NPs of
required shapes.

From the experimental point of view, several improvements of existing techniques
will be helpful to characterize the shape of the NPs. One is TEM
tomography that
allows to image an object in three dimensions, by automatically taking
a series of
pictures of the same particle at different tilt angles [\citet{Midgley2003}].
Another improvement of TEM is environmental HRTEM that is able to image
nanoparticles, with atomic lattice resolution, at various temperatures
and pressures
[\citet{Hansen2002}].

From the modeling point of view, we used marked point process to
represent the NPs
in the image, where points represent the location of NPs and marks
represent their
geometrical features. More specifically, we treated the NPs in the
image as objects,
wherein the geometrical properties of the object were largely
determined by
templates and the interaction between the objects was modeled using the area
interaction process prior. By varying the template parameters and
applying operators
such as scaling, shifting and rotation to the template, we modeled
different shapes
very realistically. In our current applications, we chose circle,
triangle, square
and ellipse as our templates. Other templates can be also constructed
in the same
framework. To solve the intractability of the posterior distribution,
we proposed a
complex Markov chain Monte Carlo (MCMC) algorithm which involves
Reversible Jump,
Metropolis--Hastings, Gibbs sampling and a Monte Carlo
Metropolis--Hastings (MCMH) for
the intractable normalizing constants in the prior. The first steps
deal with
simulating from a pseudo posterior distribution without involving the random
normalizing constant. A generalized Metropolis-within-Gibbs sampling
with a
reversible jump step is used to simulate from a pseudo posterior
distribution given
the number of objects. Additionally, a reversible jump MCMC with the
use of
birth-death and merge-split moves is invoked on moving from a state
with a different
number of objects. Finally, we simulate from the intractable
normalizing constant
posterior using Monte Carlo Metropolis--Hastings where the acceptance
ratio of the
sample taken from the pseudo posterior is estimated by simulating from
an auxiliary
variable. We reported the posterior summary statistics of the shapes
and the number
of objects in the image. We successfully applied this algorithm to real
TEM images,
outperforming convention tools aided by manual screening. Our proposed
methodology
can help practitioners to associate morphological characteristics to
physical and
chemical properties of the NPs, and in synthesizing materials that have
potential
applications in optics and medical electronics, to name a few.

\begin{appendix}
\section{Swap move}\label{appA}

Two new variables $(u_{T_j}=g_{T_j}^r,v_{T_j}=g_{T_j}^r)$ are
introduced to make it
clear that the pure parameters have a different meaning from template
to template.\vspace*{1pt} For
all the shapes, we provide a general algorithm:
Let
$\psi_j^k=(T_j^k,s_{T_j^k},\break c_{T_j^k},\theta_{T_j^k},u_{T_j^k})$\vspace*{-1pt}
denote the
current state and $\psi_j^*=(T_j^*,s_{T_j^*},c_{T_j^*},\theta
_{T_j^*},v_{T_j^*})$
the proposed state for $\psi_j^{k+1}$. The notation of the parameters
is different
from the previous sections to show the dependence of the parameters on
the model
$T_j^*$ (or template). If $T_j^k\neq T_j^*$, generate $v_{T_j^k}$ from
the prior
distribution of the $v_{T_j}$ and consider a bijection:
$(s_{T_j^*},c_{T_j^*},\theta
_{T_j^*},u_{T_j^*},v_{T_j^*})=(s_{T_j^k},c_{T_j^k},\theta
_{T_j^k},u_{T_j^k},v_{T_j^k})$.
From\vspace*{1pt} this bijection it is clear that the Jacobian is equal to identity
matrix, $J=I$, and $|J|=1$. 
In summary, the RJ-MCMC algorithm is as follows:
\begin{itemize}
\item Select model $M_{T_j^*}$ with probability $q(T_j,T_j^k)=\pi(T_j)$.
\item Generate $v_{T_j^k}$ from $\pi(v_{T_j})$.
\item
Set $(s_{T_j^*},c_{T_j^*},\theta
_{T_j^*},u_{T_j^*},v_{T_j^*})=(s_{T_j^k},c_{T_j^k},\theta
_{T_j^k},u_{T_j^k},v_{T_j^k})$.
\item Compute the M--H ratio:
\[
\alpha=\min\biggl\{1,\frac{p^*(s_{T_j^*},c_{T_j^*},\theta
_{T_j^*},v_{T_j^*}|y)\pi(T_j^k)}{p^*(s_{T_j^k},c_{T_j^k},\theta
_{T_j^k},u_{T_j^k}|y)\pi(T_j^*)}\frac{\pi(u_{T_j^*})}{\pi
(v_{T_j^k})}|J| \biggr\},
\]
where $J$ is the Jacobian.
\item Set $\psi_j^{t+1}=\psi_j^*$ with
probability $\alpha$ and
$\psi_j^{t+1}=\psi_j^{t}$ otherwise.
\end{itemize}

\section{Birth, death, split and merge moves}\label{appB}

Let $\operatorname{Pr}(\mathrm{birth})$, $\operatorname{Pr}(\mathrm{death})$,
$\operatorname{Pr}(\mathrm{split})$ and $\operatorname{Pr}(\mathrm{merge})$ be the
probabilities of
proposing a birth, death, split or a merge move, respectively.

\subsection{Birth and death pair of moves}\label{secC.1}

In the birth step a new object $\eta_{m+1}$ is proposed with a
randomly assigned
center. In this step we increase the dimension of the parameters by
$Q_{m+1}$, all
the parameters which describe the proposed object
$(\eta_{m+1},\mu_{m+1},\sigma_{m+1}^2$). All these new parameters
are sampled from
the prior distributions of the $Q_{m+1}$ parameters. The introduction
of these kind
of auxiliary variables leads again to a Jacobian equal to $1$ and the
M--H ratio is
%
\begin{equation}\label{equC.1}\quad
\min{ \biggl\{1,\frac{p^*(\eta_{m+1},\mu_{m+1},\sigma^2_{m+1},{\bolds
\bbbeta_{m},\bmu_{m},\bsigma^2_{m}}|y)}{p^*({\bolds
\bbbeta_{m},\bmu_{m},\bsigma^2_{m}}|y)\pi(\eta_{m+1},\mu
_{m+1},\sigma_{m+1}^2)}}\frac{q((m+1)\rightarrow m)}{q(m\rightarrow
(m+1))} \biggr\}.
\end{equation}

The death proposal chooses one object, $\eta_{j}$, at random and
removes it from the
configuration. The M--H ratio for this move is similar to (\ref{equ9}).

\subsection{Split and merge pair of moves}\label{secC.2}
The details for the split and merge move are more complicated than the
move types
described above. First we restrict our attention only to the case where
we merge two
neighboring objects or split one object into two neighbors. The
distance between the
two neighbors can be approximated by a function of their individual
size. When we move from one state to another, we
require that the proposed objects have equal area with the existing. In
order for
the Markov chain to be reversible we should ensure that every jump step
can be
reversed. We can improve the acceptance rate of these
moves with different proposed algorithms, for example, \citet{Al2}, but
that is beyond the scope of this paper.

To facilitate the representation, we will denote by bold characters
$\bbbeta$, $\bmu$
and $\bsigma^2$ the current state in every move and $\bbbeta_{-(\cdot)}$,
$\bmu_{-(\cdot)}$
and $\bsigma_{-(\cdot)}^2$ the current state values without the $(\cdot)$ objects.

\textit{Merge step}: Lets suppose we have two objects and that their
parameters are
$(\eta_i,\eta_j,\mu_{i},\mu_{j},\sigma^2_{i},\sigma^2_{j})$. In
the merge step, we
move to a new object with parameters
$(\eta_h,\mu_h,\sigma^2_h)=(x_h,y_h,s_h,\theta_h,T_h,g^r_h,\mu
_h,\sigma_h)$. The
equation which links the sizes of the old objects $(s_i,s_j)$ with the
new is
$s_h=\sqrt{s_i^2+s_j^2}$. Also, $x_h$ and $y_h$ are chosen to
represent the
``weighted middle'' point, taking in account the size of each object as
$(x_h,y_h)=(\frac{s_jx_j+s_ix_i}{s_i+s_j},\frac
{s_jy_j+s_iy_i}{s_i+s_j})$. All the
other parameters are chosen from one of the ``parent'' objects or at random.

In order to match the two dimensions, we introduce six auxiliary variables,
$(u_1,u_2,u_3,u_4,u_5,u_6)$, which not only would enable us to move
from state to
state but also are interpretable: $u_1=\sqrt{(y_j-y_i)^2+(x_j-x_i)^2}$
is expressing
the distance between two centers of the neighboring objects,
$u_2=\arctan((y_j-y_i)/\sqrt{(y_j-y_i)^2+(x_j-x_i)^2})$ is the angle created
from the union of the two centers $(c_1,c_2)$, $u_3=(s_i^2-s_j^2)/(s_i^2+s_j^2)$
is chosen such that $R_i=R_h\sqrt{\frac{1+u}{2}}$ and $R_j=R_h\sqrt{(1-u)/2}$,
$u_4=\theta_2$, $u_5=T_2$, $u_6=g^2_2$.

The acceptance ratio, $\alpha$, in this case is the minimum of one and
%
\begin{equation}\label{equC.2}
\frac{p^*(\eta_{h},\mu_h,\sigma^2_h,
\bbbeta_{-(i,j)},\bmu_{-(i,j)},\bsigma^2_{-(i,j)}|y)}{p^*(\eta
_{(i,j)},\mu_{(i,j)},\sigma^2_{(i,j)},\bbbeta_{-(i,j)},
\bmu_{-(i,j)},\bsigma^2_{-(i,j)}|y)} \frac{q(1\rightarrow
2)}{q(2\rightarrow1)}\frac{\prod_{i=1}^6\pi(u_{i})}{1}|J|,\hspace*{-28pt}
\end{equation}
where $|J|$ is the determinant of the Jacobian for the transformation and
$q(1\rightarrow2)$ is the split proposed probability and
$q(2\rightarrow
1)$ is the merge proposed probability.

\textit{Split step}: In the split step, we move from
($x$, $y$, $s$, $\theta$, $T$, $g^r$, $u_1$, $u_2$, $u_3$, $u_4$, $u_5$, $u_6$) to
$(x_1$, $y_1$, $x_2$, $y_2$, $s_1$, $s_2$, $\theta_1$, $\theta
_2$, $T_1$, $T_2$, $g^r_1$, $g^r_2)$.
In order to make this move possible, we introduce six proposal
distributions for the
auxiliary variables. We propose $u_1/2$ from the prior of the size
parameter, $u_2$
from the prior of rotation parameter, $u_3$ from $\operatorname{Unif}(-1,1)$,
$u_4,u_5,u_6$
from the priors of $\theta, T$ and $g^r$, respectively. In order for
this move to be
reversible, we again use the same transform that was used in the merge
step. With the same setting we can compute the M--H acceptance ratio.
\end{appendix}

\section*{Acknowledgments}

We thank Professor Faming Liang for providing the preprint of his
work on simulating from posterior distributions with doubly intractable
normalization constants. We thank all the reviewers for their useful
comments and a special thanks to the reviewer who helped us to improve
the algorithm presented in page $12$.

\begin{supplement}
\stitle{Templates and Metropolis--Hastings updates of $(\eta, \mu, \sigma)$\\}
\slink[doi]{10.1214/12-AOAS616SUPP} 
\sdatatype{.pdf}
\sfilename{aoas616\_supp.pdf}
\sdescription{Details in MCMC algorithm.}
\end{supplement}


\printaddresses

\end{document}